\documentclass[conference]{IEEEtran}
\IEEEoverridecommandlockouts
\usepackage{cite}
\usepackage{amsmath,amssymb,amsfonts,bm}
\usepackage{algorithmic}
\usepackage{graphicx}
\usepackage{subfigure}
\usepackage{textcomp}
\usepackage{xcolor}
\usepackage{csquotes}
\usepackage{ccaption}
\usepackage{verbatim}
\usepackage{tikz}
\newcommand*{\circled}[1]{\lower.7ex\hbox{\tikz\draw (0pt, 0pt)%
    circle (.5em) node {\makebox[1em][c]{\small #1}};}}

\def\BibTeX{{\rm B\kern-.05em{\sc i\kern-.025em b}\kern-.08em
    T\kern-.1667em\lower.7ex\hbox{E}\kern-.125emX}}

\IEEEoverridecommandlockouts
\IEEEpubid{978-1-7281-5416-9/19~\$31.00 \copyright~2019 IEEE}
\begin{document}


\title{Hardware Accelerator for Adversarial Attacks on Deep Learning Neural Networks}

\author{\IEEEauthorblockN{Haoqiang Guo, Lu Peng, Jian Zhang, Fang Qi, Lide Duan\textsuperscript{*}}
\IEEEauthorblockA{Louisiana State University, \textsuperscript{*}Alibaba Group\\
\{ghaoqi1, lpeng, jz, fqi1\}@lsu.edu, lide.duan@gmail.com}}

\maketitle

\begin{abstract}
Recent studies identify that Deep learning Neural Networks (DNNs) are vulnerable to subtle perturbations, which are not perceptible to human visual system but can fool the DNN models and lead to wrong outputs. A class of adversarial attack network algorithms has been proposed to generate robust physical perturbations under different circumstance. These algorithms are the first efforts to move forward secure deep learning by providing an avenue to train future defense networks, however, the intrinsic complexity of them prevents their broader usage.

In this paper, we propose the first hardware accelerator for adversarial attacks based on memristor crossbar arrays. Our design significantly improves the throughput of a visual adversarial perturbation system, which can further improve the robustness and security of future deep learning systems. Based on the algorithm uniqueness, we propose four implementations for the adversarial attack accelerator ($A^3$) to improve the throughput, energy efficiency, and  computational efficiency.  

\end{abstract}


\section{Introduction}

Recent breakthroughs in Artificial intelligence (AI) especially Deep Learning \cite{b17} have made great advances in many application fields including Autonomous Driving, Go game, and High-frequent trading in security markets, etc. However, recent studies in the computer vision area identify that Deep-learning Neural Networks (DNNs) are vulnerable to subtle perturbations to inputs which are not perceptible to human beings but can fool the DNN models and lead to wrong outputs\cite{b21}. 
Many algorithms were proposed to generate robust visual adversarial perturbations under different physical circumstances \cite{b7}\cite{b22}\cite{b25}\cite{b26}. These algorithms basically include a forward-propagation process and a error-propagation process. The former is similar to an inference procedure in regular Convolution Neural Networks (CNNs), while the latter is different from conventional CNN training process in that the inputs rather than the synaptic weights are changed in the back-propagation process. 

\begin{figure}
\centering
\includegraphics[scale=0.55]{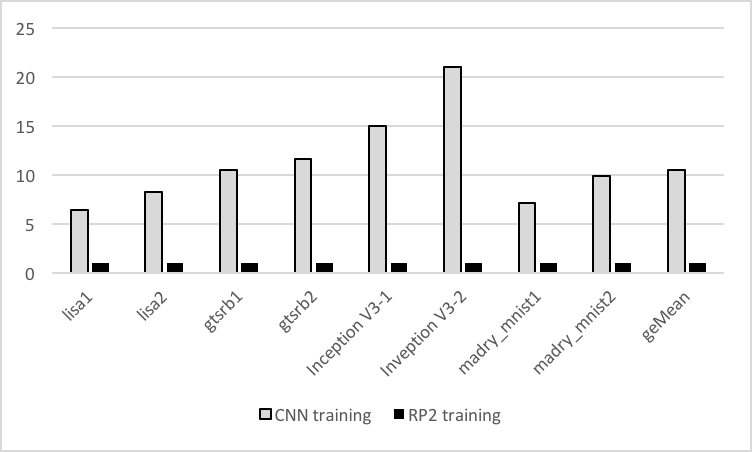}
\caption{Neuron storage analysis}
\label{memreq}
\end{figure}

The RP2 algorithm \cite{b7} and other adversarial  attack  networks (AttackNet) \cite{b22}\cite{b25}\cite{b26} are the first efforts to move toward secure deep learning by providing an avenue to train future defense networks \cite{b27}\cite{b28}\cite{b33}\cite{b34}. However, the intrinsic complexity of these algorithms prevents broader usage of them. Although recent proposals for CNN training architectures may apply to AttackNet, they were not specifically designed for the unique needs of the algorithm, resulting in low efficiency. The CNN training process consists of a Forward Propagation (FP) and a Backward Propagation (BP) for both errors and weights, while AttackNet includes the FP but a different BP for errors only. For a network with depth $L$, the duration of the neurons of layer $l$ in the buffer is $2(L-l)+1$ in CNN training process, which indicates that a large buffer is needed to minimize the gap between memory and the processor. Meanwhile, for the RP2 algorithm, the duration of neurons of any layer in the buffer is only 1. We quantitatively analyzed the neuron storage requirements of benchmarks used in this work in Fig.\ref{memreq}. The experiment results have been normalized to that of RP2 training. We can see that the overall storage requirement of the RP2 training is less than 1/10 of the CNN training. Although we take RP2 as a specific example to demonstrate our designs in this work, the optimization techniques we introduced can also be applied to other adversarial  attack  networks.

\IEEEpubidadjcol

Analog \emph{in-situ} memristor crossbar arrays have been widely used as a potential neural network accelerator because it largely reduces the energy cost of data movement. Prior research\cite{b1}\cite{b2}\cite{b4}\cite{b10} demonstrated that memristor designs outperform GPU and ASIC designs \cite{b31}\cite{b32} in both throughput and energy saving. Although Pipelayer \cite{b10} has developed a CNN training platform based on memristor crossbar arrays, directly deploying AttackNet to Pipelayer is highly inefficient \cite{b35}. First, AttackNet does not require the same amount of on-chip buffers as what CNN training does. The back propagation process of AttackNet only involves error propagation without updating weights, which can relieve the neuron storage requirement. Unnecessary on-chip buffers consume too much energy and area. Second, the utilization of memristor crossbar arrays is low. Pipelayer utilizes dual-crossbar to store weight values, which are calculated by collecting the difference between a positive crossbar subarray and a negative crossbar subarray. In this way, the on-chip crossbar utilization has been reduced by half.   

In this paper, we first identify the uniqueness of the typical visual adversarial perturbation algorithm. We analyze its dataflow and data dependence. To our knowledge, we propose the first hardware accelerator for adversarial attacks ($A^3$) using memristor crossbar arrays to significantly improve the throughput of the visual adversarial perturbation system. As a result, the robustness and security of future deep learning systems can also be improved. The main contributions of this paper can be summarized as follows. \begin{itemize}
    \item We quantitatively demonstrate that directly deploying AttackNets training on existing CNN training platforms is inefficient in both performance and energy.
    \item We explore methods of buffer reduction to increase compute engines and therefore improve the performance. 
    \item We incorporate single crossbar storage into the training process to improve the crossbar utilization and to further boost the performance.
\end{itemize}

\section{background}

\subsection{The training process of CNN and AttackNet Algorithms}
\begin{figure}[htbp]
\centering\includegraphics[scale = 0.39]{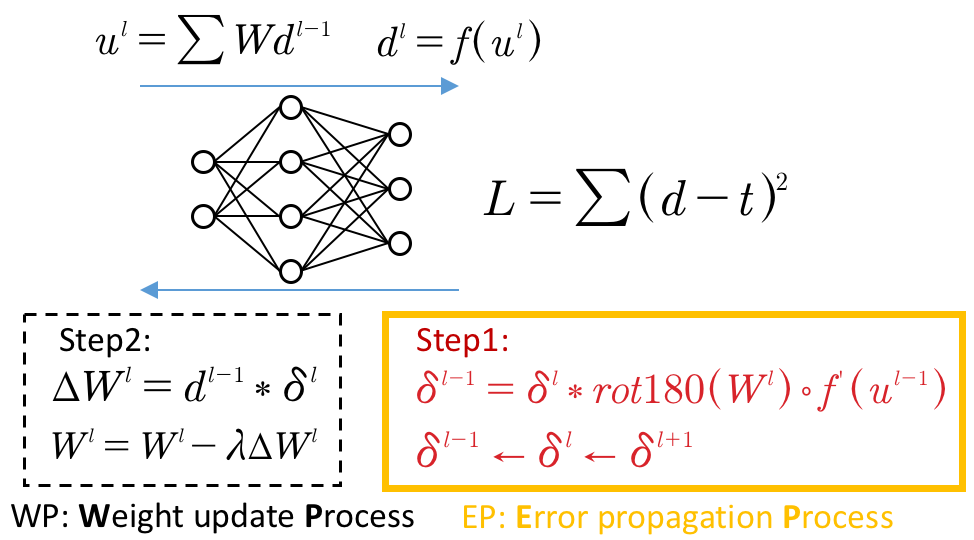}
\caption{CNN training algorithm \emph{\textbf{V.S.}} AttackNet training algorithm. $\ast$ and $\circ$ denote the convolution operation and the element-wise production respectively.}
\label{CNNt}
\end{figure}

The network architecture of both applications is cascaded, layer by layer. As shown in Fig.\ref{CNNt}, the training process of CNN can be divided into two stages, forward propagation (FP) and back propagation (BP). We denote the output of the neurons at layer $l-1$ as $d^{l-1}$. During FP, The output of the neurons at the next layer (layer $l$) can be computed by first convolving $d^{l-1}$ with the weight matrix $W^l$, then going through an activation function $f$. There are two steps in the BP of CNN. The first step is to compute errors (also called sensitivity in some literature) and propagate them backwards layer by layer. Such process is denoted as EP in this work. In the second step, the gradients of the weights at each layer $l$, $\Delta W^l$, are computed based on the error $\delta^l$ and
the output $d^{l-1}$. Once all the gradients are available, the weight matrices for all layers can be updated in parallel
using gradient descent. We refer this weight-related process as WP.

For AttackNet, its FP is roughly the same as that of the CNN. However, in contrast to CNN, whose BP includes both the 
EP and the WP processes, AttackNet's BP stage involves only EP, with the addition that it also computes the partial derivative 
of the loss function with respect to the perturbation input, in order to update the input. (The weights of the AttackNet
are not updated and thus there is no need to compute $\Delta W^l$ neither.)

\begin{figure}[htbp]
\includegraphics[scale = 0.3]{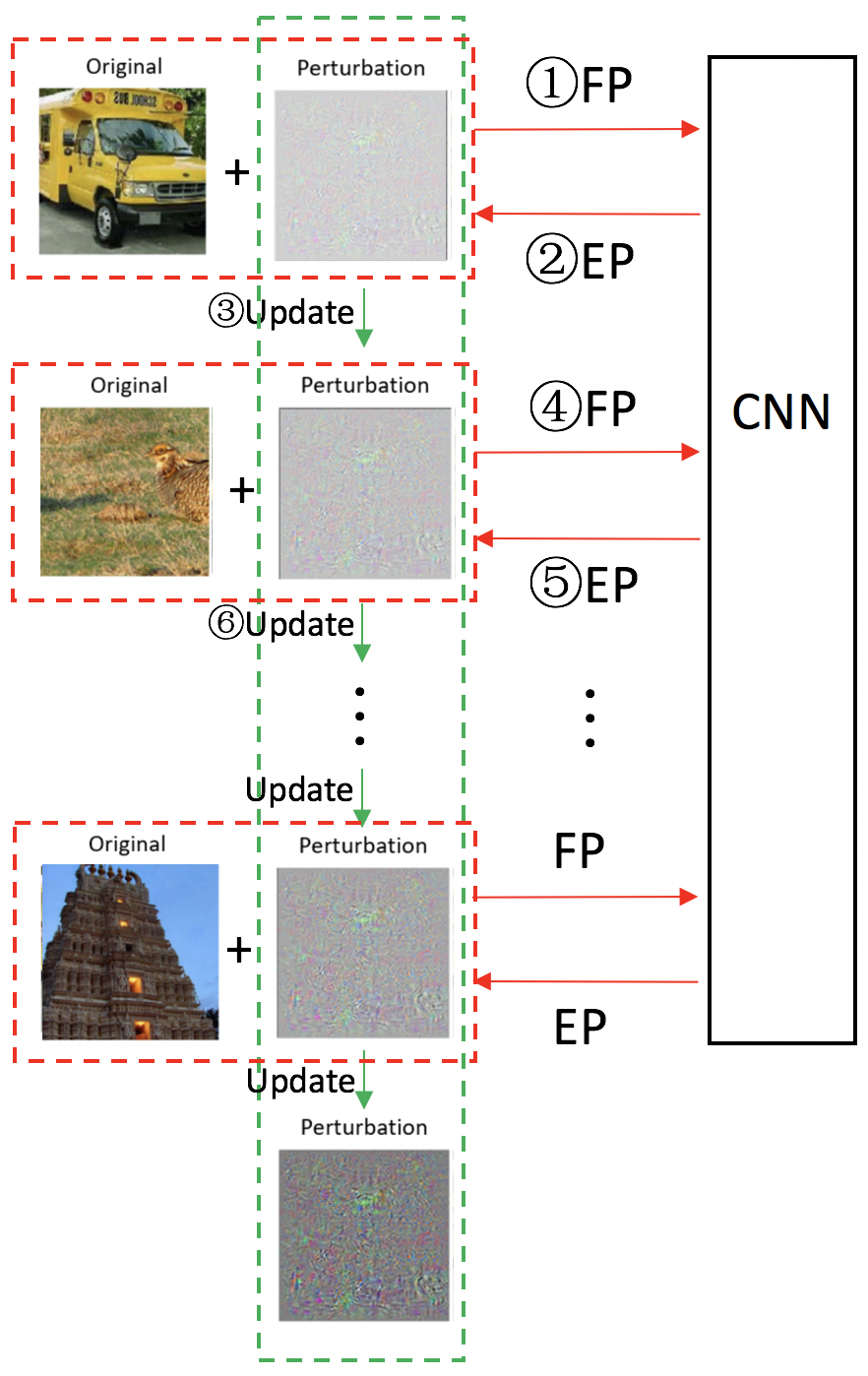}
\centering\caption{AttackNet training scheme at a high level.}
\label{CNNpg}
\end{figure}

\subsection{AttackNet Applications}
At a high level, AttackNet takes in an adversarial image, which is a combination of the original/clean image and a perturbation image (called \enquote{mask} in some literatures), to proceed the FP process. While the loss function of a regular CNN is often based on the outputs of the FP and the target labels, the loss of an AttackNet additionally incorporates the norm of the perturbation image. In BP, different from CNN training, AttackNet only executes the EP process. At the end of BP, it computes the error/derivative of the perturbation image and uses the derivative to update the perturbation (mask). This completes one training iteration. Multiple iterations are taken to train an optimal attack perturbation.

\subsection{Neural Network Acceleration in Crossbar}
There are a lot of work about neural network processing in memory. A neural network is composed of convolutional layers, pooling layers and full connected layers. All of above layers' computation can be transformed into a Matrix-Vector multiplication. For a convolutional layer or a full connected layer, the weights of it are programmed in crossbar in a matrix which is called weight matrix. For max pooling layer, a matrix of $4 \times 6$ (which is composed of 0,1,-1) is programmed in crossbar \cite{b2}. For average pooling layer, a matrix of $4 \times 6$ (which is composed of 1) is programmed in crossbar.

Generally, the resistance $1/g_{ij}$ (the crossing point at the \emph{i}th row and the \emph{j}th column) can only be positive value, while the weight value may be negative, so a \enquote*{positive} crossbar and a \enquote*{negative} crossbar are utilized to store a matrix of rational values in dual-crossbar storage situation. If a weight value $w_{ij}$ is negative, $g_{ij-}$ is larger than $g_{ij+}$ so that $g_{ij+} - g_{ij-} = w_{ij}$. If a weight value $w_{ij}$ is positive, $g_{ij+}$ is larger than $g_{ij-}$ so that $g_{ij+} - g_{ij-} = w_{ij}$. Considering the limited resistance of crossbar cells, bit slicing was used to store a single weight value, which means multi-cells in the same word line are used to store a value.

The crossbar performs a matrix-vector multiplication by imposing different voltages, which equal the components of the vector, to different word lines. The current flows along the bit line can be seen as the dot product of two vectors. The above process being performed in many bit lines can represent a matrix-vector multiplication.

\section{$A^{3}$ architecture}

In this section, we first introduce the adversarial attack accelerator ($A^{3}$) architecture, followed by the first optimization with trading off buffer storage with compute engines and the second optimization about improving the crossbar utilization. Finally, we present other associated unit designs.

\begin{figure}
\centering
\includegraphics[scale=0.8]{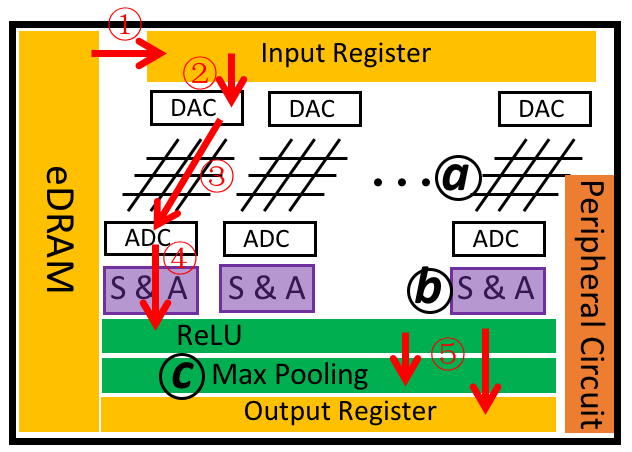}
\caption{$A^{3}$ architecture. Black circle \textcircled{$a$},\textcircled{$b$} and \textcircled{$c$} refer to crossbar, Shift\&Add unit and Max pooling unit, respectively. Red number circles refer to data flow.}
\label{arc}
\end{figure}

At a high  level, $A^{3}$ is mainly composed of storage units and functional units. The functional units are composed of crossbars, DACs, ADCs, Shift\&Add units, activation functional units, and max-pooling units. The storage units consist of eDRAM and input/output registers. We additionally deploy a peripheral circuit unit, in which the most important component is a finite state machine to control instruction flow. To support the training process, the peripheral circuit also covers a subtractor and a multiplier to calculate the errors at the end of FP. The overall architecture is shown in Fig.\ref{arc}.

The data flow of $A^3$ is shown in red number circles in Fig.\ref{arc}. During FP, inputs are loaded from eDRAM to input registers (step \textcolor{red}{\textcircled{1}}), then fed to crossbars after going through a digital-to-analog converter (DAC) (step \textcolor{red}{\textcircled{2}}). The crossbars perform the matrix-vector multiplication. The results of dot-product operations flow through an analog-to-digital converter (step \textcolor{red}{\textcircled{3}}), shift and accumulate the adjacent bit line in Shift\&Add units (step \textcolor{red}{\textcircled{4}}). Next, the dot-product results are forwarded to activation units. All the above operations are performed in a logical cycle. After the activation, there are two paths. If the layer is followed by a pooling layer, the data flows into another cycle to perform the pooling operation. At the end of this cycle, the results are stored to output registers. Otherwise, the results bypass the pooling unit and are stored to output register directly (step \textcolor{red}{\textcircled{5}}).  

\begin{figure}
\centering
\includegraphics[scale=0.4]{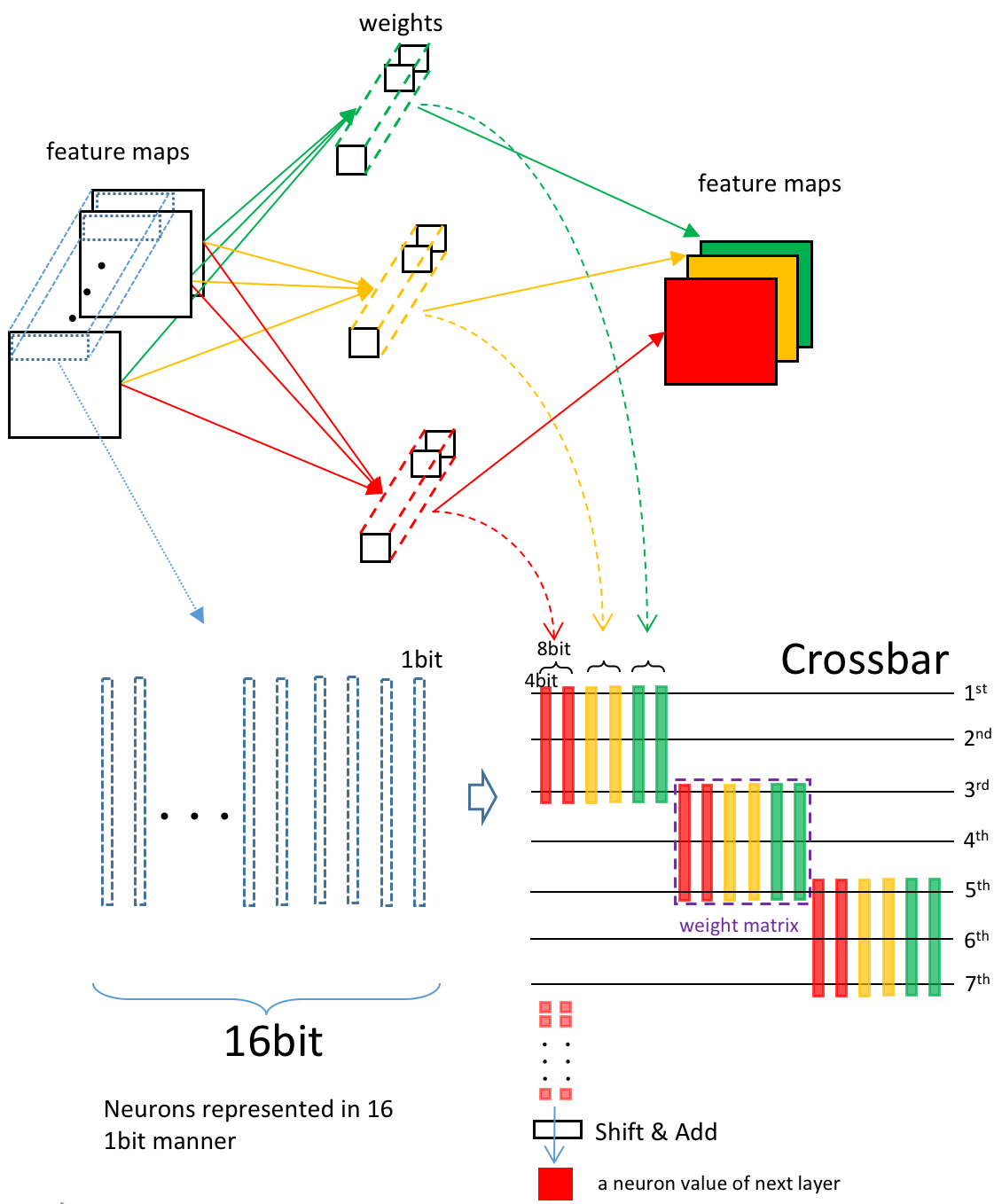}
\caption{Weight mapping to crossbars. Two cell in the same row to store a weight value, and each cell is 4-bit. The technique is called bit slicing.}
\label{map}
\end{figure}

The weights of a network layer are programmed into crossbars in the form of a matrix. The method to transform the computations of a convolutional layer can be found in cuDNN documents \cite{b8}. The strategy of mapping weights to crossbars is similar to \cite{b1}, as shown in Fig.\ref{map}. The weight matrices corresponding to the connections of the same channel (red) of the next layer are mapped to a logical bit line, followed by matrices of other channels (yellow, green) of the same layer. All the logical bit lines of the same layer make up of a logical mapping matrix. To maximize the usage of overlapping neurons of adjacent convolution operations, multiple copies of the logical mapping matrix are put after each other when programming the weight matrix to the crossbar. In $A^3$, we use two physical bit lines to represent a logical line. 
Different from traditional PIM platforms such as Pipelayer, $A^{3}$ exploits memory (eDRAM, input/output registers) optimization with the same power budget and the same area. It is unnecessary to allocate as large a buffer space as Pipelayer because $A^3$ only performs the error propagation (EP) process. Reducing memory brings power and area reduction at the same time. The power reduction or area reduction makes room for crossbars. To further increase the throughput, $A^{3}$ uses a single crossbar to store weight values, nearly doubling the utilization of crossbars. Furthermore, we redesign the Shift\&Add units and Max pooling units to support the modification. More details will be shown in the following sections.

\subsection{Trading off storage with compute engines: \bm{$A^3p$} and \bm{$A^3r$}}

As illustrated in Fig.\ref{CNNt}, CNN needs to compute the gradient of weights before updating the corresponding weight matrix. The neurons of feature maps on layer $l-1$ are involved when the system computes the gradient of weights of layer $l$, so the feature maps of each layer should be stored in the buffer for gradient computing. In pipeline cases, buffer requirement increases with the network depth. In the training process of AttackNet, weights no longer need to be updated, so neurons of feature maps of a layer can be dropped once the next layer gets its neurons. The reason lies in that the system does not compute the gradient of weights anymore. The buffer requirement of AttackNet relies on two types of data, derivatives of the ReLU function with respect to feature maps and errors of each layer. ReLU is $max(x,0)$ whose derivative is a $0$-$1$ matrix, which means the system is capable of storing elements of a matrix using a bit map rather than using a floating matrix. In this way, the required on-chip buffer can be significantly reduced.

\begin{figure}
\centering
\includegraphics[scale=0.35]{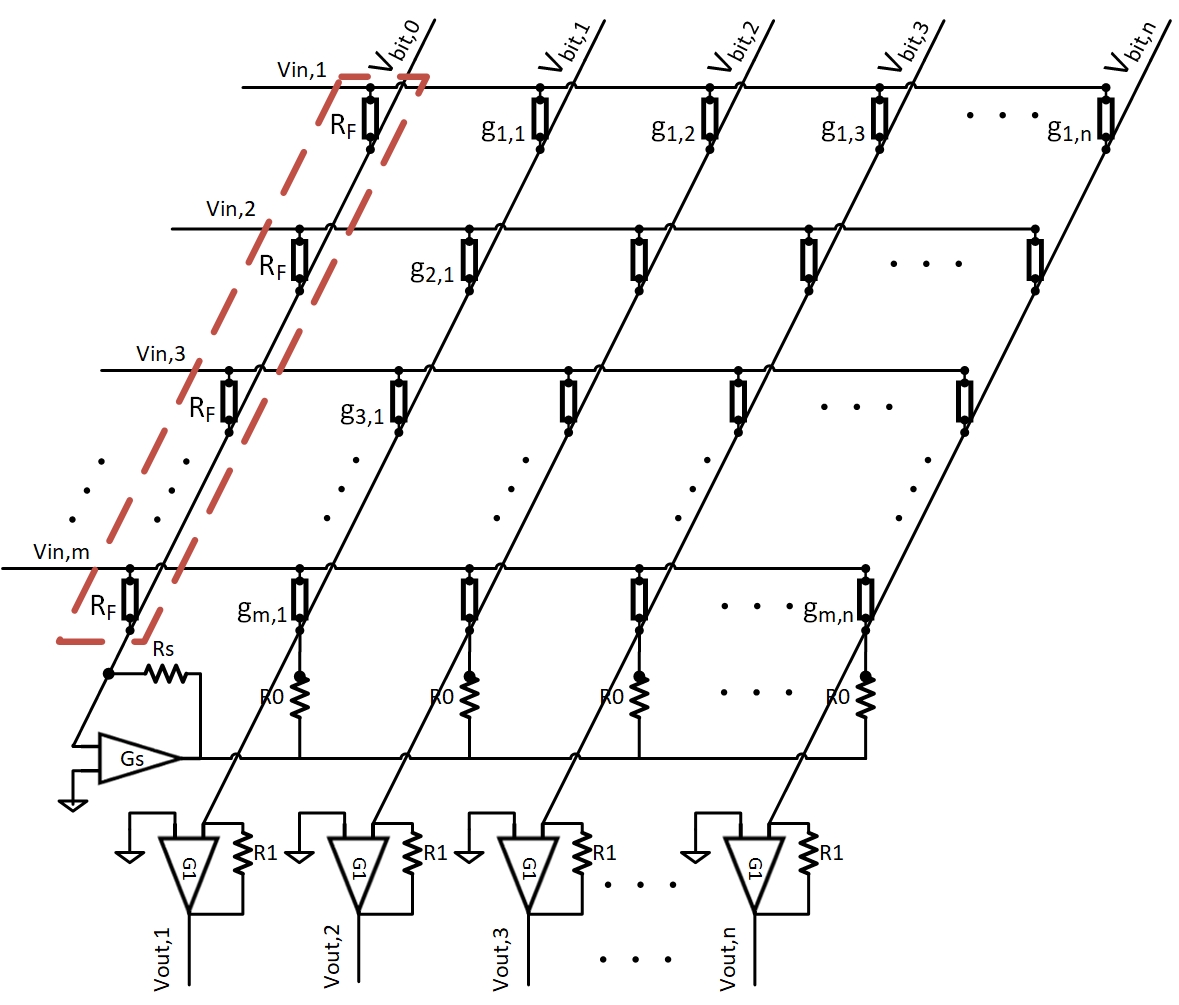}
\caption{Weight storage using single crossbar}
\label{crossbar}
\end{figure}

In this paper, we will use Pipelayer as the baseline because it is the only processing-in-memory (PIM) platform supporting CNN training. The power and area of the baseline are summarized in Table \ref{t1}. According to the observation shown in Fig.\ref{memreq}, it is unnecessary to equip $A^3$ with large buffers. Therefore, we can trade off the buffer space with more crossbars for faster computations. Assuming that the buffer being taken away accounts for \bm{$P$} and \bm{$A$} for power and area respectively, we can make use of the same power or area to add more crossbars separately. We denote the power of a crossbar and associated units as \bm{$p$}, and the area of a crossbar and associated units as \bm{$a$}. 

Under different constraints, we can have two design options. The first design \bm{$A^3p$} is to increase the number of crossbars and associated units by \bm{$P/p$}, leading to the same power budget with the baseline. The second design \bm{$A^3r$} increases the number of crossbars and associated units by \bm{$A/a$}, thus resulting in the same area space to the baseline. As we will see in the methodology section, \bm{$A^3r$} will incur power overhead, but the performance efficiency of it is slightly higher than that of baseline for some benchmarks.


\begin{figure}
\centering
\includegraphics[scale=0.6]{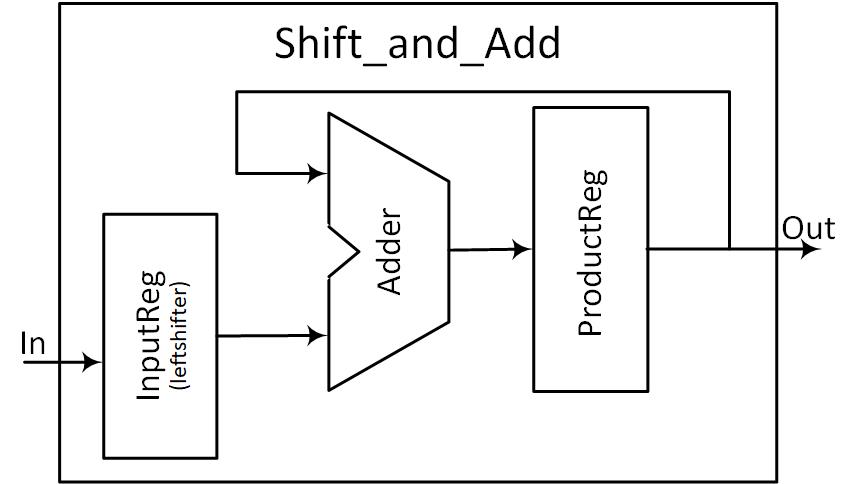}
\caption{Shift\&Add unit design}
\label{sa}
\end{figure}

\subsection{Improving crossbar utilization: \bm{$A^3px$} and \bm{$A^3rx$}}

\begin{figure}
\centering
\includegraphics[scale=0.28]{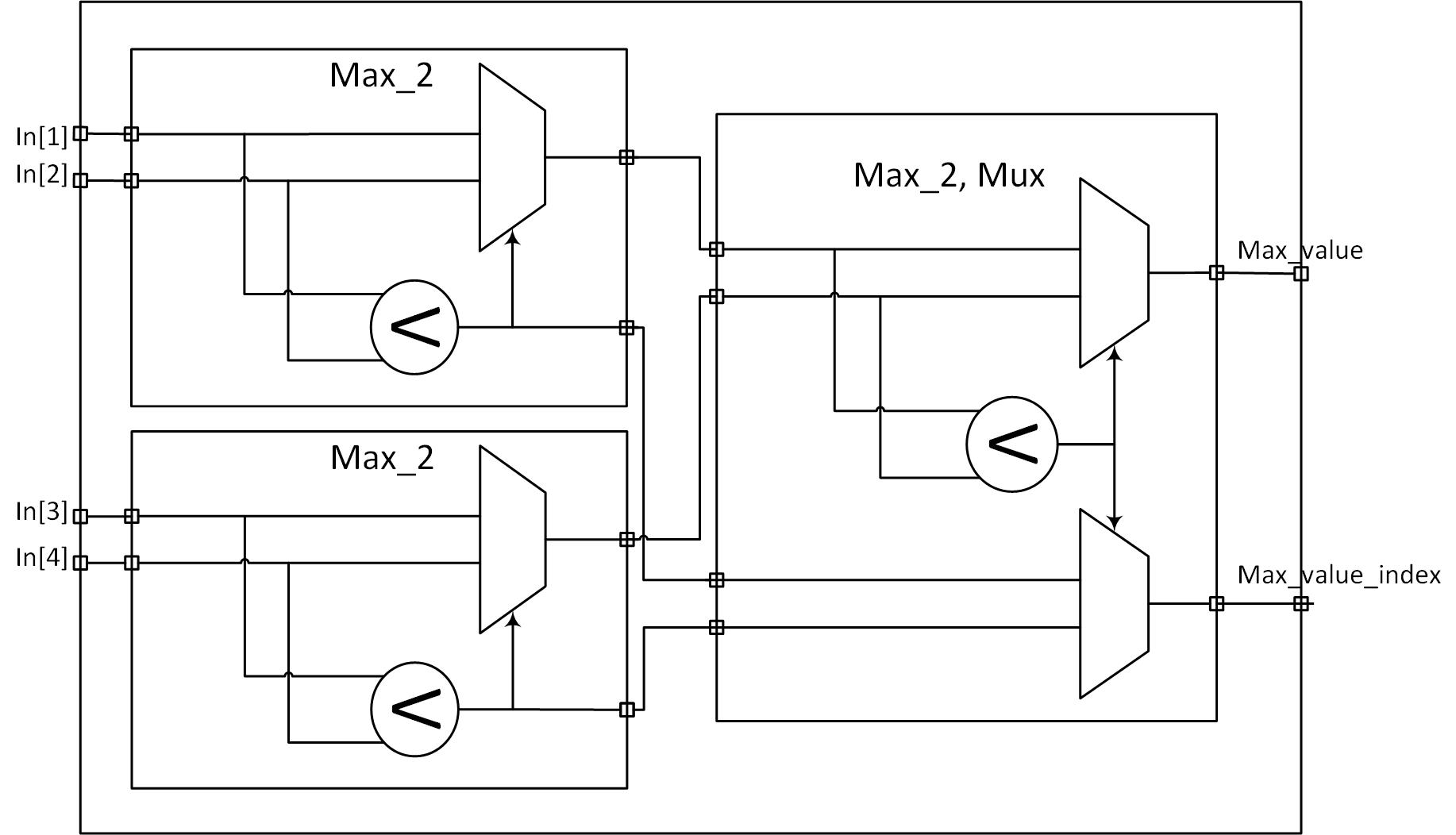}
\caption{Max pooling unit design}
\label{mp}
\end{figure}

Traditionally, dual-crossbar is used to store weight matrices \cite{b1}\cite{b2}\cite{b10}. In this manner, the difference between the two crossbars equals the actual weight values. The drawback of this method is that only half of crossbars on-chip are used to perform dot-production.    

As illustrated in Fig.\ref{crossbar}, the crossbar of $A^3$ adds a column of constant-term circuits $1/R_F$ on the left of the custom crossbar. The memristor's conductance at the crossing point of the \emph{i}th row and \emph{j}th column is $g_{i,j}$. $V_{in,m}$ is the voltage applied to the \emph{m}th row. According to Kirchhoff's current law, the current along the \emph{k}th bit line is 
\begin{equation}\nonumber
    V_{out,k} = -\left [ \sum_{s=1}^{m}(R_1\cdot g_{s,k}-\frac{R_1}{R_F})V_{in,s} \right]
\end{equation}
This value can be directly fed to the analog-to-digital unit instead of going through a subtractor. The overhead incurred by the constant-term $1/R_{F}$ and the OP amp is negligible \cite{b9}. With this minor modification, we can store weights using a single crossbar instead of two crossbars and improve resource utilization significantly.

The single crossbar optimization can be incorporated with the proposed \bm{$A^3p$} and \bm{$A^3r$} designs to further improve the crossbar utilization. We refer the combined techniques as \bm{$A^3px$} and \bm{$A^3rx$} respectively.

\subsection{Shift\&Add Unit and Max pooling Unit Design}

The multiplication of two binary numbers comes down to calculating partial products. The Shift\&Add unit is shown in Fig.\ref{sa}. The most significant 4 bits enter the unit, are shifted left, then being accumulated with the least significant 4 bits. The output of the adder is the product of the original two binary numbers.    

The max pooling unit design is presented in Fig.\ref{mp}. Overall, the unit is implemented in a tree-like manner with depth equaling two. In the first stage, it takes in four inputs and compares the two pair of numbers in parallel. The outputs of the first stage not only include the comparison result of each comparator but also cover the input indices of the comparator results. Both the indices are fed to a multiplexer whose output is determined by another comparator in the second stage. Finally, the max pooling unit outputs the maximum value of the four inputs, and the index corresponds to that maximum value. 

\textbf{Differences with PipeLayer:} \bm{$A^3$} explores buffer optimization due to the uniqueness of AttackNet algorithms, and redesigns functional units to accommodate weight storage in single crossbars. All optimization techniques enable \bm{$A^3$} to outperform Pipelayer in throughput and energy efficiency compared to deploying AttackNets on PipeLayer.

\section{Pipeline Analysis}

\begin{figure}[htbp]
\includegraphics[scale = 0.35]{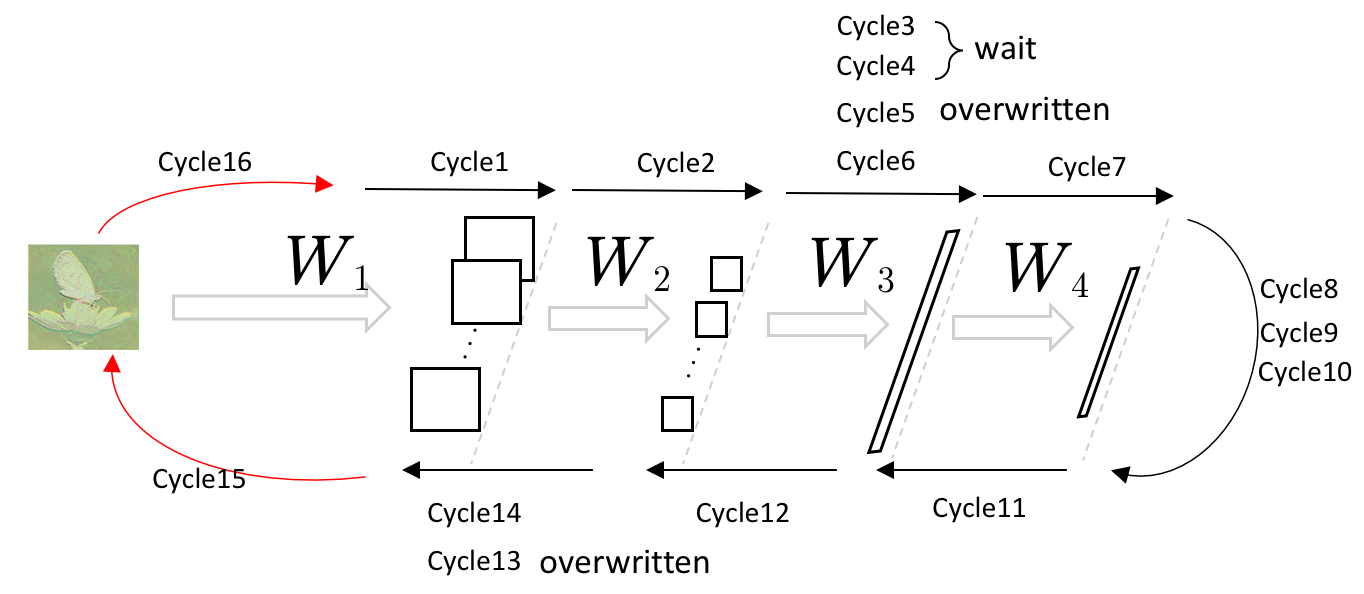}
\centering\caption{An AttackNet training example: two convolutional layers and two fully connected layers. The batch size is 3.}
\label{cy1}
\end{figure}

\begin{figure*}
\centering
\includegraphics[scale=0.35]{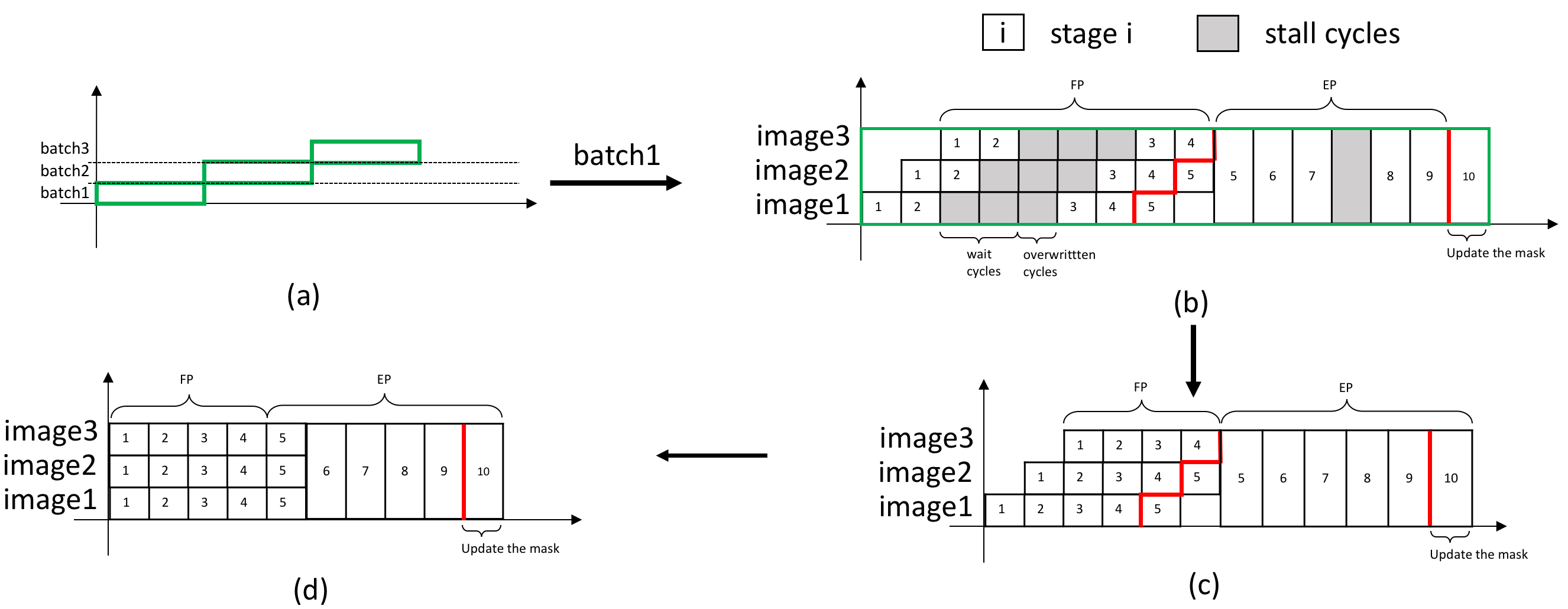}
\caption{Pipeline analysis for the network example.}
\label{pip}
\end{figure*}

In this section, we take a network of four layers as an example to illustrate how crossbar counts affect performance. The AttackNet example in Fig.\ref{cy1} is composed of two convolutional layers ($L_1$,$L_2$) and two fully connected layers ($L_3$,$L_4$).The weight matrix of each layer is $W_1$, $W_2$, $W_4$ and $W_4$. Each layer accounts one cycle when its weight matrix is available. Otherwise, it should wait for its weight matrix, which can overwrite other weight matrices if and only if they have been used by the last image in the same batch.

\begin{figure}[htbp]
\includegraphics[scale = 0.35]{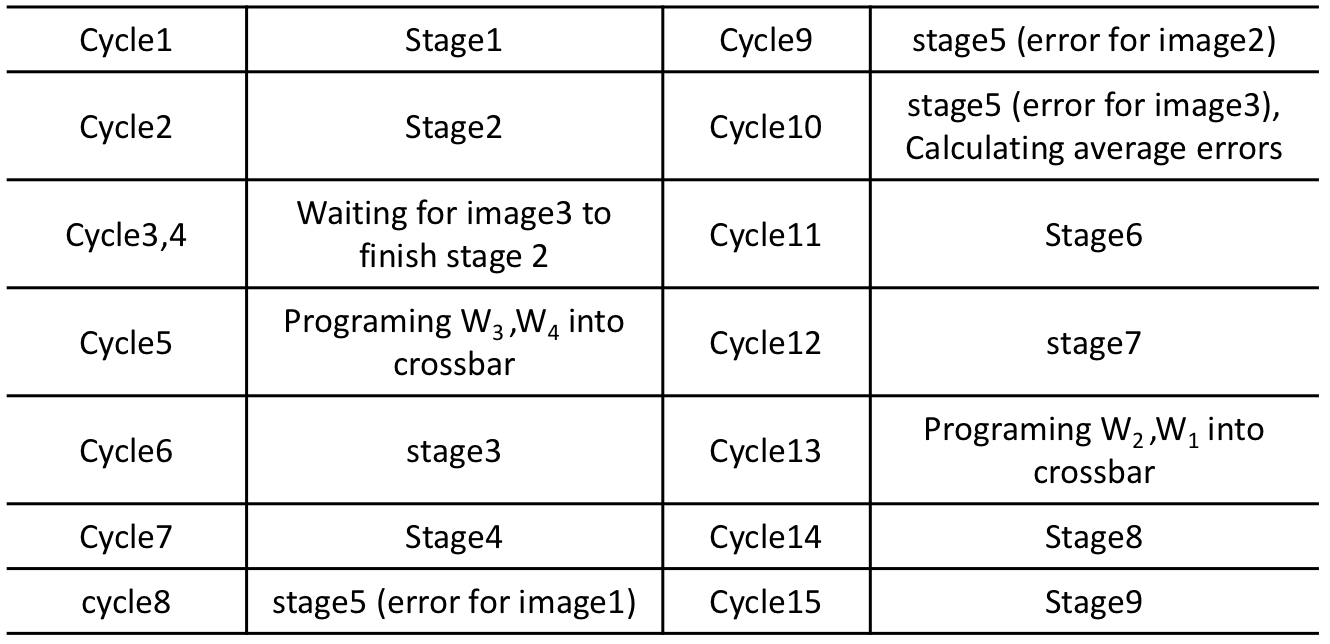}
\centering\caption{Pipeline stages for the AttackNet training example.}
\label{cy2}
\end{figure}

Assuming a memristor crossbar based CNN training platform is equipped with $C$ crossbars, we can only pre-program $W_1$ and $W_2$ into these $C$ crossbars. In this case, Fig.\ref{pip} shows the pipeline details when batch size equals three. Fig.\ref{pip}(b) focuses on image1 of the first batch. The system processes $L_1$ (corresponding to stage1) in the 1st cycle, followed by stage2 that proceeds the computation of $L_2$ in the 2nd cycle. The system should wait for $W_1$ and $W_2$ being used by image3, meanwhile stall image1 in the following two cycles and image2 in the 4th cycle, then is capable of programming $W_3$ and $W_4$ into crossbars in the 5th cycle. After that, the system performs the computation of the two fully connected layers, $L_3$ and $L_4$, in the 6th cycle and the 7th cycle respectively. In the 8th cycle and the 9th cycle, the system computes the error of $L_4$ for image1 and image2 respectively. During the 10th cycle, the system computes the error of $L_4$ for image3 and accumulate errors for all images in this batch. Note that the error computation of $L_4$ needs target labels instead of a weight matrix. The system can process $L_3$ and $L_2$ in the following two cycles (the 11th and 12th cycle) because $W_4$ and $W_3$ are already in crossbars. After the 12th cycle, the system needs to wait for an extra cycle to program $W_2$ and $W_1$ into crossbars. Finally, the system computes the error of $input$ in the 15th cycle, then updates the mask of inputs in the next cycle. All the above processes are illustrated in Fig.\ref{pip}(b) and summarized in Fig.\ref{cy2}.

As crossbar counts increase, \bm{$A^3$} significantly decreases the overwritten times, thus reducing the processing time of a batch. The Fig.\ref{pip}(c) shows the pipeline when \bm{$A^3$} can store the whole network parameters (the weight matrix of all layers). More crossbars delete stall cycles, thus reducing the processing time of one batch of inputs. Further increasing crossbars enables more than one set of the network parameters to be programmed into crossbars. Fig.\ref{pip}(d) shows the pipeline when \bm{$A^3$} can store three copies of the whole network parameters. As we can see, the pipeline of multiple input sets is similar to that of the super-scalar processor.

\section{Methodology}

\begin{table}[]
\resizebox{90mm}{12mm}{
\tiny
\begin{tabular}{|c|c|c|c|c|c|c|}
\hline
\multicolumn{7}{|c|}{Baseline (Pipelayer)}                                                                                   \\ \hline
\multicolumn{1}{|l|}{} & Size    & Power(w)    & Area(mm\textasciicircum{}2) & \#count & Total Power & Total Area \\ \hline
eDRAM buffer           & 32MB    & 4.49 & 16.364                      & 1       & 4.49 & 16.364     \\ \hline
Output Register        & 128KB   & 0.04 & 0.175                   & 1       & 0.037 & 0.175  \\ \hline
Input Register         & 128KB   & 0.037 & 0.1752                   & 1       & 0.037 & 0.175  \\ \hline
crossbar                   & 128*128 & 0.0003      & 0.000025                    & 16128   & 4.84      & 0.4     \\ \hline
DAC                    & 1*128   & 0.0005      & 0.00002125                  & 16128   & 8.064       & 0.34272    \\ \hline
ADC                    & 8bits   & 0.002       & 0.0012                      & 16128   & 32.256      & 19.3536    \\ \hline
SUM                    &         &             &                             &         & 49.72 & 36.814 \\ \hline
\end{tabular}}
\caption{Baseline (Pipelayer) Configuration}\label{t1}
\end{table}

\begin{table}[]
\resizebox{90mm}{12mm}{
\tiny
\begin{tabular}{|c|c|c|c|c|c|c|}
\hline
\multicolumn{7}{|c|}{$A^{3}p$}                                                                                   \\ \hline
\multicolumn{1}{|l|}{} & Size    & Power(w)    & Area(mm\textasciicircum{}2) & \#count & Total Power & Total Area \\ \hline
eDRAM buffer           & 2MB    & 1.36 & 2.45                      & 1       & 1.36 & 2.45     \\ \hline
Output Register        & 16KB   & 0.01 & 0.015                   & 1       & 0.01 & 0.01  \\ \hline
Input Register         & 16KB   & 0.01 & 0.015                   & 1       & 0.01 & 0.01  \\ \hline
crossbar                   & 128*128 & 0.0003      & 0.000025                    & 17265   & 5.1795      & 0.43     \\ \hline
DAC                    & 1*128   & 0.0005      & 0.00002125                  & 17265   & 8.63       & 0.3669    \\ \hline
ADC                    & 8bits   & 0.002       & 0.0012                      & 17265   & 34.53      & 20.718    \\ \hline
SUM                    &         &             &                             &         &  49.719 & 23.992 \\ \hline
\end{tabular}}
\caption{$A^{3}p$ Configuration}\label{t2}
\end{table}

\begin{table}[]
\resizebox{90mm}{12mm}{
\tiny
\begin{tabular}{|c|c|c|c|c|c|c|}
\hline
\multicolumn{7}{|c|}{$A^{3}r$}                                                                                   \\ \hline
\multicolumn{1}{|l|}{} & Size    & Power(w)    & Area(mm\textasciicircum{}2) & \#count & Total Power & Total Area \\ \hline
eDRAM buffer           & 2MB    & 1.36 & 2.45                      & 1       & 1.36 & 2.45     \\ \hline
Output Register        & 16KB   & 0.01 & 0.015                   & 1       & 0.01 & 0.01  \\ \hline
Input Register         & 16KB   & 0.01 & 0.015                   & 1       & 0.01 & 0.01  \\ \hline
crossbar                   & 128*128 & 0.0003      & 0.000025                    & 27553   & 8.27      & 0.689     \\ \hline
DAC                    & 1*128   & 0.0005      & 0.00002125                  & 27553   & 13.78       & 0.586    \\ \hline
ADC                    & 8bits   & 0.002       & 0.0012                      & 27553   & 55.11      & 33.064   \\ \hline
SUM                    &         &             &                             &         &  78.53 & 36.813 \\ \hline
\end{tabular}}
\caption{$A^{3}r$ Configuration}\label{t3}
\end{table}

\begin{figure}[htbp]
\includegraphics[scale = 0.4]{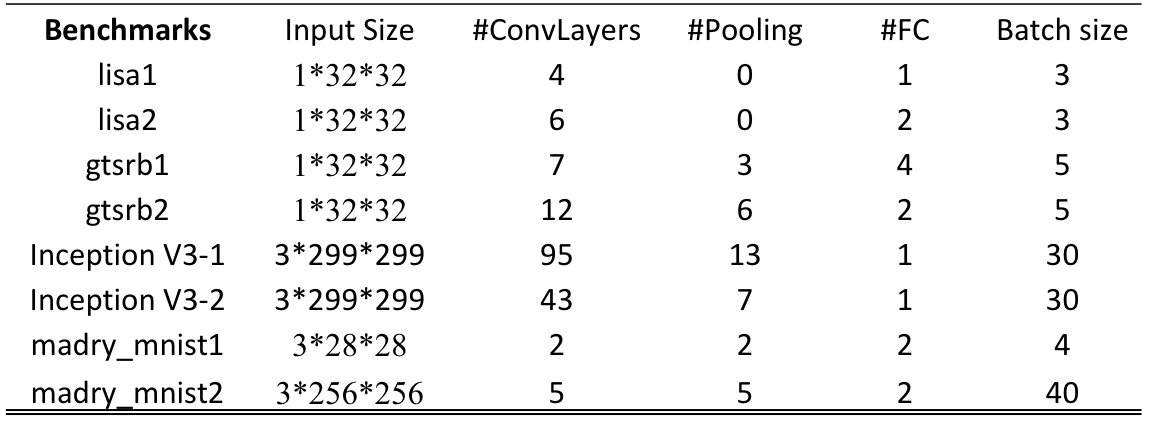}
\centering\caption{Benchmarks and the configurations.}
\label{bench}
\end{figure}

\textbf{Power and Area Model:} In this work, we use CACTI 7.0\cite{b11} at 32 nm to model the power and area of the SRAM buffer (input/output registers). The consumption of power and area of eDRAM are measured using Destiny\cite{b12} at 32 nm. The area and energy for memristor-based crossbar are adapted from \cite{b13}\cite{b14}. The area and power of DAC and ADC are modeled from the analysis \cite{b15} and ISAAC\cite{b4}. For simplicity, we use a 1-bit DAC. The power and area of the  Shift\&Add unit, the Max pooling unit are negligible when compared with other units. We developed the baseline accelerator following Pipelayer as described in \cite{b10} with configurations listed in table \ref{t1}. $A^3p$ and $A^3r$ are shown in table \ref{t2} and table \ref{t3} respectively.

\textbf{Performance Model:} To simulate the process, we developed an in-house simulator to model the training process of adversarial attack algorithms, and to estimate the throughput performance. The cycle time in $A^3$ is 50.88ns, which is consistent with Pipelayer\cite{b10}. The metrics we use are power efficiency (PE, the number of 16-bit operations performed per watt, $GOPs/W$) and computational efficiency (CE, the number of 16-bit operations performed per second per $mm^2$, $GOPs/T \times mm^2$), both of which are defined in ISAAC \cite{b4}.

\textbf{Benchmarks:} The benchmarks we used in this paper are selected from from \cite{b7} \cite{b16}. The benchmark configurations are listed in Fig.\ref{bench}. lisa1, gtsrb1, Inception v3-1 and madry\_mnist1 are the same as that in the original literature. lisa2, gtsrb2, Inception v3-2 and madry\_mnist2 are modified from lisa1, gtsrb1, Inception v3-1 and madry\_mnist1 to make the test networks' architecture more diverse.

\section{Experiment Results}

\begin{figure}
\centering
\subfigure[Performance improvement on the same power budget]{
\begin{minipage}[b]{1\linewidth}\centering
\includegraphics[width=0.9\linewidth]{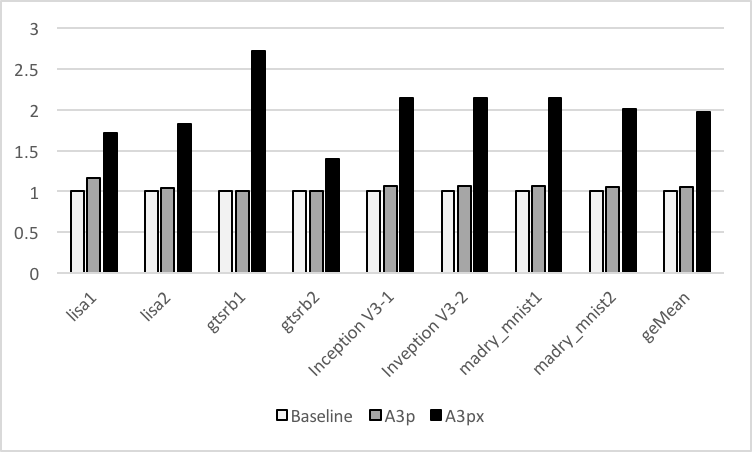}
\end{minipage}\label{memopt1}}

\subfigure[Performance improvement on the same area budget]{
\begin{minipage}[b]{1\linewidth}\centering
\includegraphics[width=0.9\linewidth, height=4.5cm]{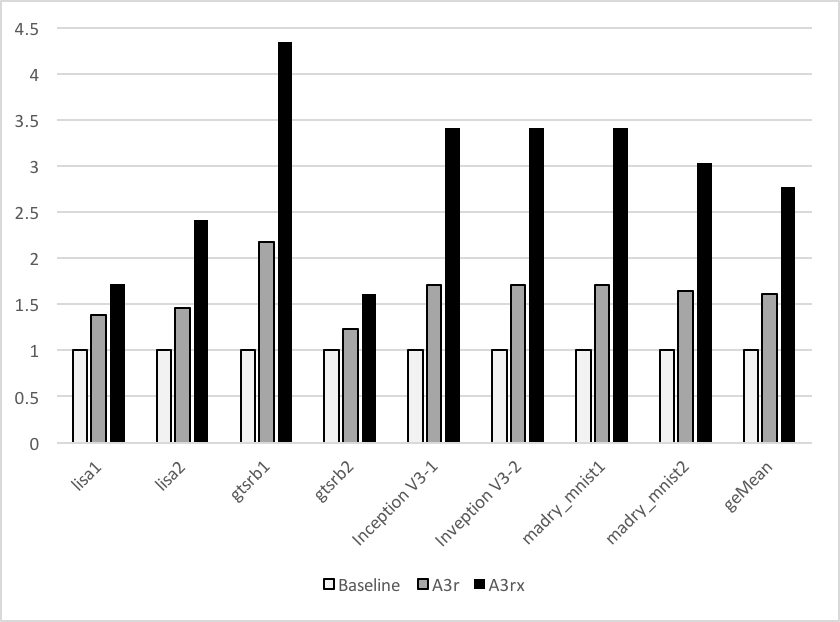}
\end{minipage}\label{memopt2}}
\caption{Speedup results over the baseline. All results are normalized to the baseline.}\label{memopt}
\end{figure}

According to the previous discussion, \bm{$A^3p$} and \bm{$A^3r$} have the same power and the same area to the baseline respectively. To further increase the throughput, we apply single crossbar storage to both designs. All the performance results are shown in Fig.\ref{memopt}. From Fig.\ref{memopt1}, reducing the number of buffers to make room for crossbars can bring a little performance improvement under the same power budget due to the limited power saving brought by buffer reduction. In contrast, single crossbar storage significantly increases performance. The results of keeping the same area budget are presented in Fig.\ref{memopt2}. The speedup of all benchmarks on \bm{$A^3r$} is larger than that of \bm{$A^3p$}. The area reduction brought by buffer reduction enables more crossbars to be added to \bm{$A^3$}. As can be seen in table \ref{t2} and table \ref{t3}, \bm{$A^3r$} owns more crossbar than \bm{$A^3p$}.      

\begin{figure}[htbp]
\centering\includegraphics[scale = 0.6]{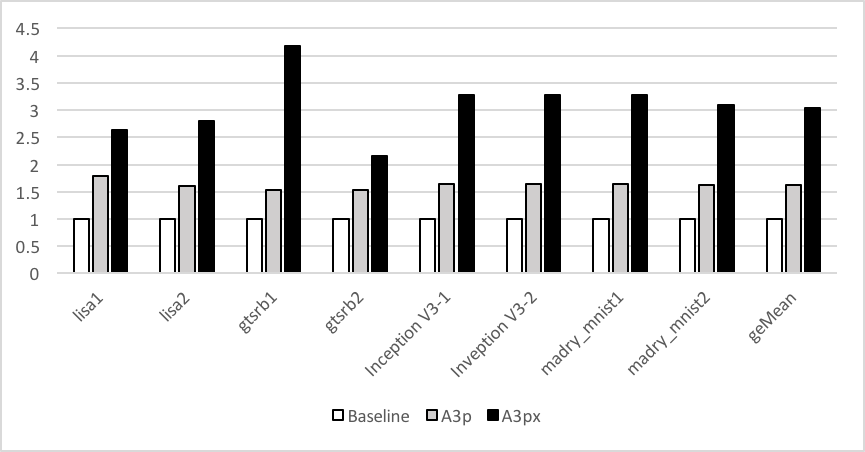}
\caption{Computational efficiency improvement on $A^3p$}
\label{ce}
\end{figure}

Since the power of \bm{$A^3p$} is the same as that of the baseline, we only compare the computational efficiency results in Fig.\ref{ce}. Compared with \ref{memopt1}, CE speedups of all benchmarks on \bm{$A^3p$} are more obvious. The reason can be found in table \ref{t1} and \ref{t2}. Keeping the same power budget, the area of \bm{$A^3p$} is smaller than that of the baseline. 

\begin{figure}[htbp]
\centering\includegraphics[scale = 0.6]{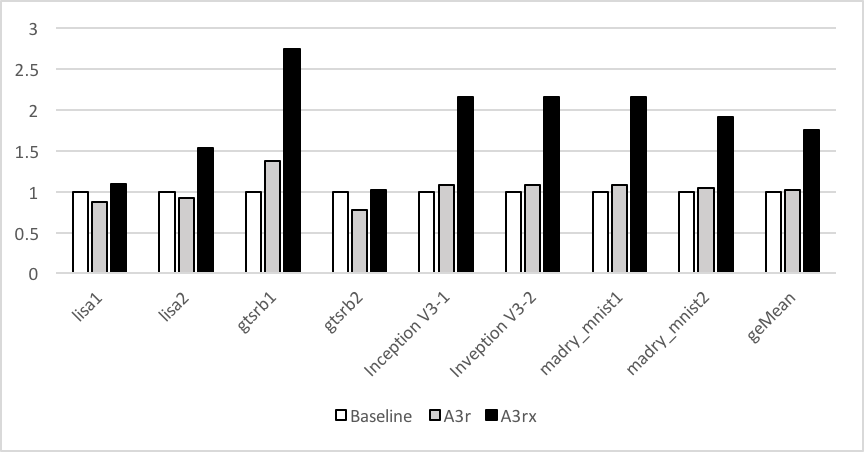}
\caption{Power efficiency improvement on $A^3r$}
\label{pe}
\end{figure}

We only compare the power efficiency of all benchmarks on \bm{$A^3r$} in Fig.\ref{pe} because the area of \bm{$A^3r$} is the same as that of the  baseline. Though increasing  crossbars in \bm{$A^3r$} incurred power overheads, the geometric mean of the PE is better than that of the baseline. However, some benchmarks' PE on \bm{$A^r$} are slightly worse than that on the baseline. 

\begin{figure}
\centering
\subfigure[The percentage of overwritten on the same power budget]{
\begin{minipage}[b]{1\linewidth}\centering
\includegraphics[width=0.95\linewidth]{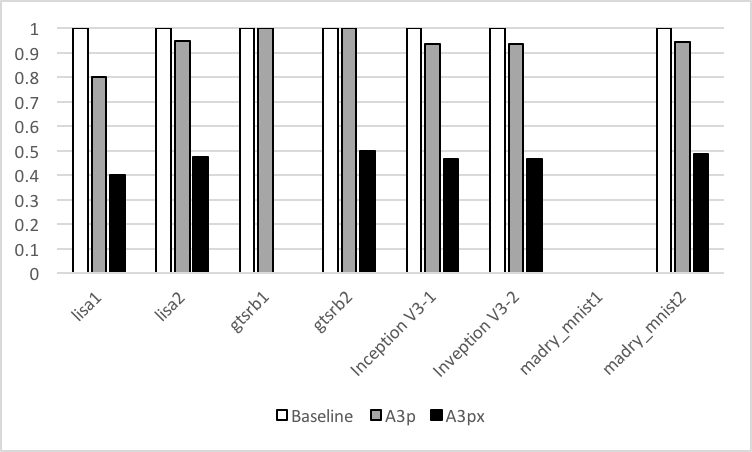}
\end{minipage}\label{ow1}}

\subfigure[The percentage of overwritten on the same area budget]{
\begin{minipage}[b]{1\linewidth}\centering
\includegraphics[width=0.95\linewidth]{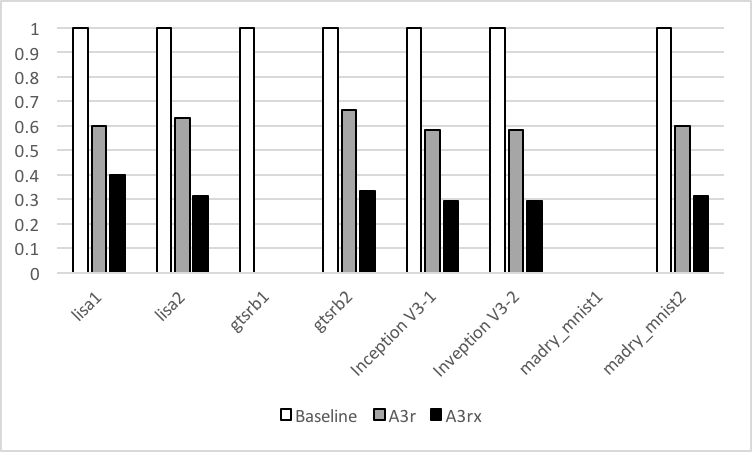}
\end{minipage}\label{ow2}}
\caption{Weight overwritten proportions vary with optimization. All the results are normalized to the baseline.}\label{ow}
\end{figure}

Fig.\ref{ow} shows the results of weight overwritten times of all benchmarks. The overwritten times of \bm{$A^3p$}, \bm{$A^3px$}, \bm{$A^3r$} and \bm{$A^3rx$} are normalized to that of the baseline. Relatively, \bm{$A^3px$} and \bm{$A^3r$} reduce the numbers of weight overwritten much more than \bm{$A^3p$} and \bm{$A^3r$} do. Madry\_minst1 does not have any weight overwritten in all cases. Gtsrb1 does not need overwrite weights on \bm{$A^3r$}, \bm{$A^3px$}, and \bm{$A^3rx$}. Overall, we can observe that reducing overwritten times can bring performance benefits for all benchmarks. This figure is largely inverse to the speedups shown in Fig. \ref{memopt}.

\begin{figure}[htbp]
\includegraphics[scale = 0.35]{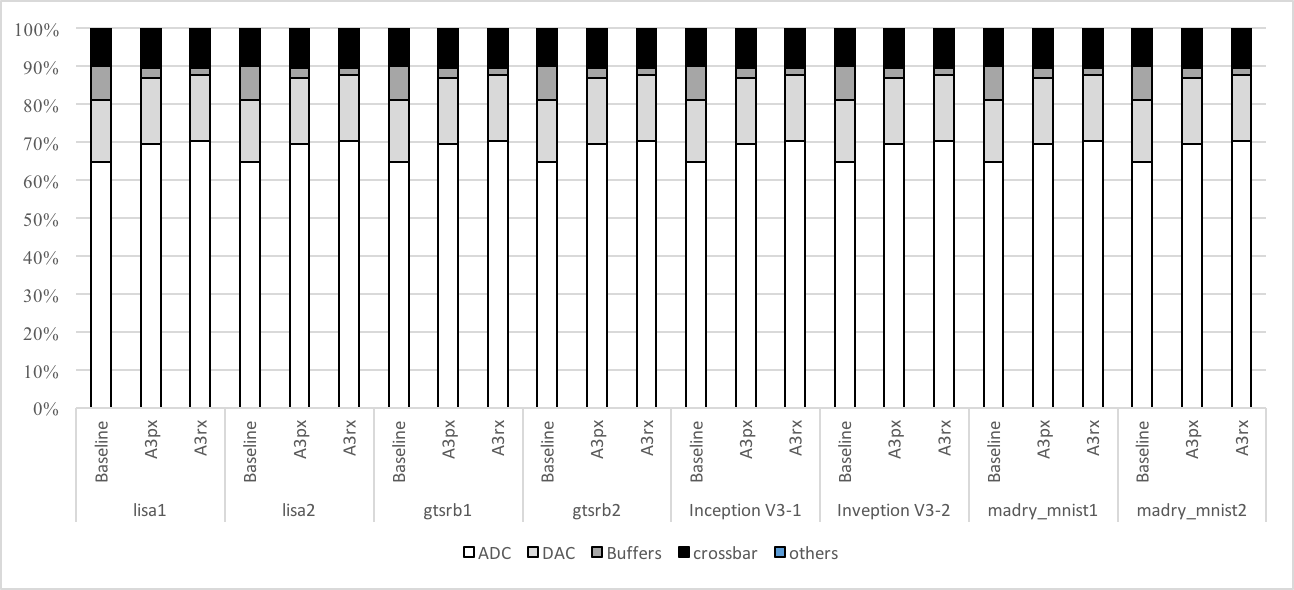}
\centering\caption{Energy Breakdown}
\label{energy}
\end{figure}

Fig.\ref{energy} provides the breakdown of the energy consumption by different components. The total energy consumption is divided into five parts, ADC energy, DAC energy, Buffer energy, crossbar energy and others. From Fig.\ref{energy}, we can see clearly the proportion of ADC consumption and DAC consumption in \bm{$A^3px$} and \bm{$A^3rx$} are larger than that in \bm{$A^3$}. The reason is that both \bm{$A^3px$} and \bm{$A^3rx$} own more crossbar than \bm{$A^3$}. The number of ADC and DAC are in proportional to crossbar numbers.

\textbf{Accuracy and convergence discussion:} AttackNet aim to misclassifying inputs, thus generating a mask for training a robust deep learning network, so they are less sensitive to accuracy than CNN training. The accuracy of AttackNet should take convergence into account, which means results should be evaluated by pairing the iteration with the corresponding misclassifying rate. The bit-widths of neurons and weights in this work are constant with ISAAC, and have no effects on AttackNet according to our observations.

\section{Conclusion}
Adversarial Attacks to CNNs are an emerging topic in the deep neural networks area. In this paper, we propose  the  first  hardware  accelerator for adversarial  attacks  based  on memristor crossbar  arrays. Compared to conventional CNN training architectures, the proposed design significantly improves performance and energy efficiency. 


\end{document}